\input harvmac
\overfullrule=0pt
\font\authorfont=cmcsc10 \ifx\answ\bigans\else scaled\magstep1\fi
{\divide\baselineskip by 4
\multiply\baselineskip by 3
\def\prenomat{\matrix{\hbox{hep-th/9606199}&\cr \qquad\hbox{SWAT/122}&\cr
\qquad\hbox{DTP/96/56}&\cr}}
\Title{$\prenomat$}{\vbox{\centerline{ Multi-Instanton Check
of the Relation} 
\vskip3pt
\centerline{Between the Prepotential ${\cal F}$ and the Modulus $u$}
\vskip3pt
\centerline{in $N=2$ SUSY Yang-Mills Theory}}}
\centerline{\authorfont Nicholas Dorey}
\bigskip
\centerline{\sl Physics Department, University College of Swansea}
\centerline{\sl Swansea SA2$\,$8PP UK $\quad$ \tt n.dorey@swansea.ac.uk}
\bigskip
\centerline{\authorfont Valentin V. Khoze}
\bigskip
\centerline{\sl Department of Physics, Centre for Particle Theory, 
University of Durham}
\centerline{\sl Durham DH1$\,$3LE UK $\quad$ \tt valya.khoze@durham.ac.uk}
\bigskip
\centerline{and}
\bigskip
\centerline{\authorfont Michael P. Mattis}
\bigskip
\centerline{\sl Theoretical Division T-8, Los Alamos National Laboratory}
\centerline{\sl Los Alamos, NM 87545 USA$\quad$ \tt mattis@pion.lanl.gov}
\vskip .3in
\def\hf{{\textstyle{1\over2}}}
\def\quarter{{\textstyle{1\over4}}}
\noindent
By examining   multi-instantons in
$N=2$ supersymmetric $SU(2)$ gauge theory, we derive, on very
general grounds, and to all orders in the instanton number,
a relationship between the prepotential ${\cal F}(\Phi)$,
and the  coordinate
on the quantum moduli space $u=\langle\Tr \Phi^2\rangle$.
This relation was previously obtained by Matone in the context of 
the explicit Seiberg-Witten  low-energy solution of the model.
Our findings can be viewed as a multi-instanton check of the proposed 
exact results in supersymmetric gauge theory.

\vskip .1in
\Date{\bf June 1996. } 
\vfil\break
}

\lref\Gates{J. Gates, Nucl.~Phys.~B238 (1984) 349.}
\lref\FT{F. Fucito and G. Travaglini, 
{\it Instanton calculus and nonperturbative relations
 in $N=2$ supersymmetric gauge theories}, 
 ROM2F-96-32, hep-th/9605215.}
\lref\Bilal{See Sec.~6  of
A. Bilal, {\it Duality in $N=2$ SUSY $SU(2)$ Yang-Mills theory:
A pedagogical introduction to the work of Seiberg and Witten}, 
LPTENS-95-53, hep-th/9601007.}
\lref\dkmMO{N. Dorey, V. Khoze and M.P. Mattis, 
{\it Multi-instanton calculus in $N=2$ supersymmetric gauge theory},
hep-th/9603136, to appear in Phys. Rev. D.}
\lref\wessbagger{J. Wess and J. Bagger, {\it Supersymmetry and supergravity}, 
Princeton University Press, 1992.} 
\lref\SWone{N. Seiberg and E. Witten, 
{\it Electric-magnetic duality, monopole
condensation, and confinement in $N=2$ supersymmetric Yang-Mills theory}, 
Nucl. Phys. B426 (1994) 19, (E) B430 (1994) 485  hep-th/9407087}
\lref\KLTtwo{A. Klemm, W. Lerche and S. Theisen, 
{\it Nonperturbative effective actions of 
$N=2$ supersymmetric gauge theories}, 
CERN-TH/95-104, hep-th/9505150.}
\lref\MAtone{
M. Matone, {\it Instantons and recursion relations in $N=2$ SUSY gauge theory}
 Phys. Lett. B357 (1995) 342,    hep-th/9506102.  }
\lref\BMSTY{For more recent work along these lines see for example: \hfil\break
G. Bonelli and M. Matone, 
    Phys. Rev. Lett. 76 (1996) 4107, hep-th/9602174; \hfil\break
            J. Sonnenschein, S. Theisen and S. Yankielowicz,
            Phys. Lett. 367B (1996) 145, hep-th/9510129.}        

\lref\FPone{ D. Finnell and P. Pouliot,
{\it Instanton calculations versus exact results in 4 dimensional 
SUSY gauge theories},
Nucl. Phys. B453 (95) 225, hep-th/9503115. }
\lref\Sone{ N. Seiberg, Phys. Lett. B206 (1988) 75. }
\lref\ADHMone{ M. F. Atiyah, V. G. Drinfeld, N. J. Hitchin and
Yu. I. Manin, Phys. Lett. A65 (1978) 185. }
\lref\Oone{ H. Osborn, Ann. Phys. 135 (1981) 373. }
\lref\CGTone{ E. Corrigan, P. Goddard and S. Templeton,
Nucl. Phys. B151 (1979) 93; \hfil\break
   E. Corrigan, D. Fairlie, P. Goddard and S. Templeton,
    Nucl. Phys. B140 (1978) 31.}
\def\bose{{\rm bose}}

\def\inst{{\rm inst}}

\def\fermi{{\rm fermi}}

\def\dalpha{{\dot\alpha}}
\def\dbeta{{\dot\beta}}
\def\dgamma{{\dot\gamma}}
\def\ddelta{{\dot\delta}}

\def\Im{{\rm Im}}
\def\sst{\scriptscriptstyle}

\def\vsd{v^{\sst\rm SD}}

\def\Phibar{\bar\Phi}
\def\F{{\cal F}}
\def\G{{\cal G}}

\def\susy{supersymmetry}

\def\ADHM{{\scriptscriptstyle\rm ADHM}}
\def\cl{{\,\rm cl}}
\def\lambdabar{\bar\lambda}

\def\psibar{\bar\psi}
\def\sqrtwo{\sqrt{2}\,}

\def\susic{supersymmetric}

\def\zero{{\scriptscriptstyle(0)}}
\def\new{{\scriptscriptstyle\rm new}}

\def\uA{\,\lower 1.2ex\hbox{$\sim$}\mkern-13.5mu A}
\def\uX{\,\lower 1.2ex\hbox{$\sim$}\mkern-13.5mu X}
\def\uD{\,\lower 1.2ex\hbox{$\sim$}\mkern-13.5mu {\rm D}}

\def\uF{\,\lower 1.2ex\hbox{$\sim$}\mkern-13.5mu F}
\def\uW{\,\lower 1.2ex\hbox{$\sim$}\mkern-13.5mu W}
\def\uWbar{\,\lower 1.2ex\hbox{$\sim$}\mkern-13.5mu {\overline W}}

\def\uV{\,\lower 1.2ex\hbox{$\sim$}\mkern-13.5mu V}
\def\uv{\,\lower 1.0ex\hbox{$\scriptstyle\sim$}\mkern-11.0mu v}
\def\uPsi{\,\lower 1.2ex\hbox{$\sim$}\mkern-13.5mu \Psi}
\def\uPhi{\,\lower 1.2ex\hbox{$\sim$}\mkern-13.5mu \Phi}
\def\uchi{\,\lower 1.5ex\hbox{$\sim$}\mkern-13.5mu \chi}
\def\Psibar{\bar\Psi}
\def\uPsibar{\,\lower 1.2ex\hbox{$\sim$}\mkern-13.5mu \Psibar}
\def\upsi{\,\lower 1.5ex\hbox{$\sim$}\mkern-13.5mu \psi}
\def\psibar{\bar\psi}
\def\upsibar{\,\lower 1.5ex\hbox{$\sim$}\mkern-13.5mu \psibar}
\def\upsibarzero{\,\lower 1.5ex\hbox{$\sim$}\mkern-13.5mu \psibar^\zero}
\def\ulambda{\,\lower 1.2ex\hbox{$\sim$}\mkern-13.5mu \lambda}
\def\ulambdabar{\,\lower 1.2ex\hbox{$\sim$}\mkern-13.5mu \lambdabar}
\def\ulambdabarzero{\,\lower 1.2ex\hbox{$\sim$}\mkern-13.5mu \lambdabar^\zero}
\def\ulambdabarnew{\,\lower 1.2ex\hbox{$\sim$}\mkern-13.5mu \lambdabar^\new}
\def\D{{\cal D}}

\def\Dslash{\,\,{\raise.15ex\hbox{/}\mkern-12mu \D}}
\def\Dbarslash{\,\,{\raise.15ex\hbox{/}\mkern-12mu {\bar\D}}}
\def\delslash{\,\,{\raise.15ex\hbox{/}\mkern-9mu \partial}}
\def\delbarslash{\,\,{\raise.15ex\hbox{/}\mkern-9mu {\bar\partial}}}
\def\L{{\cal L}}
\def\hf{{\textstyle{1\over2}}}
\def\quarter{{\textstyle{1\over4}}}

\def\uAcl{\,\lower 1.2ex\hbox{$\sim$}\mkern-13.5mu A^{}_{\cl}}
\def\uAbarcl{\,\lower 1.2ex\hbox{$\sim$}\mkern-13.5mu A_{\cl}^\dagger}

\def\Leff{\L_{\rm eff}}

\leftline{\bf 1. \authorfont Introduction.}

The concept of electric-magnetic duality has led to remarkable recent
progress in understanding the low energy physics of certain (3+1) 
dimensional supersymmetric gauge theories. In this note we focus on
pure $N=2$ \susic\ $SU(2)$ gauge theory, the exact low-energy limit of
which has been given by Seiberg and Witten \SWone. As sketched below,
their solution consists of explicit expressions for the vevs
$a(u)$ and $a_D(u)$ 
of the Higgs field $A$ and its dual field $A_D$, as functions
of the parameter $u$ 
\eqn\udef{u \ \equiv \langle \hf\, A^a A^a \rangle \ .}
Alternatively, Matone \MAtone\ has recast these as  predictions
for the $N=2$ prepotential $\F(a)$ and the parameter $u\equiv \G(a)$ as
functions of the vev $a$; explicitly, one finds \refs{\MAtone,\BMSTY}
\eqna\npr
$$\eqalignno{&\G (a) \ = \ \pi i \ 
\bigl(\F (a) \ - \ \hf a \,\F'(a)\bigr) \ , &\npr a
\cr
&\big(4\Lambda^4\,-\,\G^2\big)\G''(a)\ +\ 
\textstyle{{1\over 4}}a\,\big(\G'(a)
\big)^3\ =\
0\ .&\npr b}$$

These equations are naturally Taylor
expanded in the RG-invariant scale parameter $\Lambda^4,$
the $n$th power of which captures the contribution of the $n$-instanton
sector. Thus instantons provide an important 
means of checking the Seiberg-Witten solution \npr{} 
directly---without  appealing to duality, nor to arguments about the
number and nature of the singularities in the analytic $u$ plane. 
That Eq.~\npr{b} holds at the 1-instanton and 2-instanton level 
(with $\G$ eliminated
in favor of $\F$ as per Eq. (2a)) has
been verified in Refs.~\FPone\ and \dkmMO, respectively. Subsequently, using
the methods of Refs.~\refs{\FPone,\dkmMO},  the authors of Ref.~\FT\ have
verified Eq.~\npr{a} as well, through the 2-instanton level. In this note,
we point out that Eq.~\npr{a} is in fact  built into the instanton
calculus, and holds automatically at the $n$-instanton level, for all $n$.

\leftline{\bf2. \authorfont Review}

The particle content of pure $N=2$ \susic\ $SU(2)$ gauge theory consists, in
$N=1$ language, of a gauge multiplet $W^a_\alpha=(v^a_m, \,  \lambda^a)$ 
coupled to a complex chiral matter multiplet $\Phi^a=(A^a, \,  \psi^a)$ which 
transform in the adjoint representation of $SU(2)$, $a=1,2,3$. 
Here $v^a_m$ is the gauge field, $A^a$ is the Higgs field, Weyl 
fermions $\lambda^a$ and $\psi^a$ are the gaugino and Higgsino.
Classically, the adjoint scalar $A^a$ can have an arbitrary complex 
vev, $(0,0,a)$, breaking $SU(2)$ down to $U(1)$. 
$N=2$ supersymmetry remains unbroken, and protects the flat direction in 
$a$ from being lifted by quantum corrections.
Different values of $a$  span a one complex dimensional
family of theories, known as the quantum moduli space.
  
The isospin component of the fields that is aligned with the vev $(0,0,a)$
remains massless, whereas
the remaining two components acquire a mass $M_W=\sqrtwo |a|.$
For length scales $x\gg1/M_W$ the massless modes can be described by Wilsonian
effective action 
 in terms of an $N=1$ photon superfield
$W_\alpha=( v_m, \, \lambda)$ and chiral superfield $\Phi=( A, \, \psi).$
$N=2$ \susy\ restricts the terms in the effective action
with not more than two derivatives or four fermions to the following
form  \refs{\Sone,\Gates,\SWone}:
\eqn\Leffdef{\Leff\ =\ {1\over4\pi}\,\Im\left[\int d^4\theta\F'(\Phi)
\Phibar+\int d^2\theta\,\hf\F''(\Phi)W^\alpha W_\alpha\,\right]\  .}
Thus $\Leff$ is specified by a single object, the holomorphic 
prepotential $\F$.
Its second derivative is just the running complexified 
coupling $\tau (a)$ \SWone:  
\eqn\tauequals{\F''(a) 
\ = \ \tau (a) \ = \ {4 \pi i \over g^2 (a)} + 
   {\vartheta (a) \over 2 \pi} \ ,}
where $\vartheta$ is the effective theta-parameter.
The first derivative  $\F' (\Phi)$ 
defines the superfield $\Phi_D$ dual to $\Phi$. This
is the local field of the  dual magnetic description 
of the low-energy theory \SWone. It has vev
\eqn\ad{a_D = \langle \Phi_D \rangle = \langle  A_D \rangle=\F'(a)\ .}

The  vevs $a$ and  $a_D$ provide
alternative local parametrizations of the moduli space
of the theory. However, neither
 $a$ nor $a_D$ is a
good global coordinate on the quantum moduli space; 
 instead they are traded for a gauge-invariant 
 parameter $u$, Eq.~\udef,
which is. By determining the behavior of $a(u)$ and $a_D(u)$ in the
vicinity of their hypothesized singularities, Seiberg and Witten
were able to reconstruct them exactly \SWone:
\eqna\swaad
$$\eqalignno{
a \ &= \ {\sqrt{2} \over \pi} \ \int_{-2\Lambda^2}^{2\Lambda^2}
{ dx \sqrt{x-u} \over \sqrt{x^2-4\Lambda^4}} \ ,&\swaad a
\cr
a_D \ &= \ {\sqrt{2} \over \pi} \ \int_{2\Lambda^2}^{u}
{ dx \sqrt{x-u} \over \sqrt{x^2- 4\Lambda^4}} \ .&\swaad b }$$
 Here $\Lambda$
is the dynamically generated scale of the effective $U(1)$
theory which can be matched \FPone\ to the dynamical scale
of the microscopic $SU(2)$ theory, $\Lambda_{\sst\rm PV}$,
computed in the Pauli-Villars regularization scheme---the 
natural scheme for doing instanton calculations.
These expressions are singular at 
$u \to \pm 2 \Lambda^2$ and $u \to \infty$.
Physically, the singularities at $+2\Lambda^2$ and $-2\Lambda^2$
correspond, respectively,  to the monopole and the dyon 
becoming massless, while the singularity
at $\infty$ is a perturbative one-loop effect and follows from
the  asymptotic freedom of the original theory.

In order to extract the prepotential from Eq.~\swaad{}, one formally
inverts Eq.~\swaad{a},
$u={\cal G}(a)$, and then uses Eqs.~\ad\ and 
\swaad{b} to obtain:
\eqn\ddF{\F''(a) \ = \ { a'_D(u) \over a'(u)} \, \Big|_{u={\cal G}(a)}  \ .} 
This procedure yields the following expansions
for $u$ and $\F\,$:
\eqn\GSWdef{u \ \equiv \ \G(a)\ =
\ {\cal G}_{\rm clas}(a)\,+\,{\cal G}_{\rm inst}(a)
\ =\
{1\over 2} a^2 \ + \ 
\sum_{n=1}^\infty
{\cal G}_n\,\left({\Lambda\over a}\right)^{4n} a^2\ ,}
and  
\eqn\FSWdef{\F(a)\ =
\ {\cal F}_{\rm pert}(a)\,+\,{\cal F}_{\rm inst}(a)
\ =\
{i\over2\pi}
a^2\,\log{2 a^2\over e^3\Lambda^2}\ -\ {i\over\pi}
\sum_{n=1}^\infty
{\cal F}_n\,\left({\Lambda\over a}\right)^{4n} a^2\ .}
That the expansion for $\F$
 has this general form had been known for some time 
\Sone$\,$ with the $n$th term in the series  
being
an $n\hbox{-}$instanton effect. The new
information in the Seiberg-Witten solution is the numerical
value 
\refs{\KLTtwo, \MAtone} of each of the coefficients ${\cal F}_{n}$; in
particular,\foot{Numerical values of ${\cal F}_{n}$ depend on the
normalization of $a$ and the prescription for $\Lambda$.
In our conventions \dkmMO, 
the ${\cal F}_{n}$ are those of Ref.~\KLTtwo\ times a factor of $2^{6n-2}\,$.}
${\cal F}_1=1/2$, with the higher ${\cal F}_n$'s 
(${\cal F}_2=5/16\,,$ 
${\cal F}_3={3 \over 4}$, ${\cal F}_4={1469 \over 2^9}$,  
${\cal F}_5={4471 \over 40}$, ... )
being determined by a recursion relation implied by Eqs.~\npr{}
\MAtone.   This constitutes a set 
of highly non-trivial predictions for all
multi-instanton contributions to the low-energy physics in this theory.

{}From hereon in we focus on the relation \npr{a}. Differentiating both
sides with respect to $u$ gives the following constancy condition on the
Wronskian $W(a,a_D)\,=\,a(u)a'_D(u)-a_D(u)a'(u)$:
\eqn\wronskian{1\ =\ -{i\pi\over2}\,W(a,a_D)\ .}
Up to two constants (an integration constant, and the multiplicative
factor in Eq.~\wronskian, both of which are fixed by examining
the asymptotically free large-$|u|$ regime \MAtone), Eq.~\npr{a} is
therefore tantamount to the condition $dW/du=0.$
This, in turn, implies that $a$ and $a_D$ satisfy a
 common linear homogeneous second-order differential equation
with a vanishing first-derivative term.
For the
specific Seiberg-Witten solution \swaad{} this equation has in
fact been found, and reads \refs{\MAtone,\KLTtwo}
\eqn\diffeq{\big[(4\Lambda^2-u^2){d^2\over du^2}\,-\,\quarter\big]\,a
\ =\ 
\big[(4\Lambda^2-u^2){d^2\over du^2}\,-\,\quarter\big]\,a_D\ =\ 0\ .}
But it is clear that Eq.~\npr{a}, unlike Eq.~\npr{b}, is more
general than the specific  solution \swaad{}. As reviewed in
Ref.~\Bilal, this linear ODE condition is in fact the natural paradigm
for understanding the monodromy conditions \SWone\ on $a$ and $a_D.$

\leftline{\bf3. \authorfont Multi-Instanton Check of Eq.~\npr{a}.}

Expanding Eq.~\npr{a} in the instanton number $n$ gives the following
simple rewrite \MAtone:
\eqn\GnFn{\G_n  \ = \ 2n \F_n \ .}
This relation has recently been checked for $n=1$ and $n=2$
 in Ref.~\FT. In what follows we give the proof of Eq.~\GnFn\ for all instanton
numbers.

To understand the multi-instanton
contributions to $\cal G$ and  $\cal F$, we first need to review some general
features of the collective coordinate measure.
In bosonic pure gauge theory the general $n$-instanton solution
of Atiyah, Drinfeld, Hitchin and Manin (ADHM) \ADHMone\ 
contains $8n$ collective coordinates. 
In $N=2$ \susic\ gauge theory, these are augmented by
$8n$ fermionic collective coordinates which parametrize
the gaugino and higgsino adjoint fermion zero modes \CGTone.
Denoting the $8n$ unconstrained bosonic and fermionic coordinates as $X_{i}$
and $\chi_{i}$, respectively, one may express 
the measure for $n$ ADHM instantons \Oone\
in $N=2$ \susic\ Yang-Mills theory as follows (see Sec.~7.5 of \dkmMO):
\eqn\measureone{
\int\, d\mu_n\ =\ {1\over{\cal S}_n}\int \,
\left(\prod_{i=1}^{8n} dX_{i}d\chi_{i}\right) 
\big(\,J_\bose/J_\fermi\,\big)^{1/2}\,\exp(-S_{n\hbox{-}\inst})\ .
 }
Here $J_\bose$ ($J_\fermi$) is the Jacobian
for the bosonic (fermionic) collective coordinates,
while the  super-multi-instanton action $S_{n\hbox{-}\inst}$ was calculated 
in Sec.~7.4 of Ref.~\dkmMO. 
An important simplification in a supersymmetric theory is that there are no
additional small-fluctuations 't Hooft determinants to be calculated,
as the non-zero eigenvalues for bosonic and fermionic excitations cancel 
identically. Finally, as explained  in \refs{\Oone,\dkmMO},
to obtain the correct normalization of the
measure,  one must divide out the relevant symmetry factor ${\cal S}_{n}$. 

In truth, the construction of  ${\cal S}_n$, $J_\bose$ and $J_\fermi$
depends on one's ability to isolate $8n$ independent bosonic collective
coordinates from the unfortunately overcomplete set of ADHM 
variables---currently an unsolved problem for $n\ge3.$ However, these
issues are entirely irrelevant for  present purposes; this is because
 Eq.~\GnFn\ is diagonal in instanton number (unlike Matone's
 recursion relation \MAtone\ that follows from Eq.~\npr{b}, which
connects different $n$ sectors).

As explained in detail in Ref.~\dkmMO,
most of the zero modes of the vevless super-multi-instanton
are  lifted by the vev of the adjoint Higgs field through the Yukawa
terms in the action.
We shall find it convenient to single out from the measure  $ d\mu_n$
the integrations over the eight unlifted modes:
global translations, $d^4 x_0 $,
and the supersymmetric collective coordinates for gaugino and
higgsino,  $d^2 \xi_1$ and $d^2 \xi_2$,
\eqn\measuretwo{
\int\, d\mu_n\ =\ \int\, d^4 x_0 \  d^2 \xi_1 \ d^2 \xi_2 \ \int\, 
d\tilde\mu_n  \ . }

The general $n$-instanton contribution to the functional integral
for the quantum modulus $u$  is simply
\eqn\Ginst{ \G(a) \ 
\equiv u \ \equiv \ \langle {\hf}\, A^a  A^a \rangle  \ 
= \
\hf \, \int\, d\mu_n \ A^a_{\rm inh} \ A^a_{\rm inh} \ .}
Here $A^a_{\rm inh}$ is the solution of the inhomogeneous 
Euler-Lagrange equation,
\eqn\inhH{\D^2\uA\ =\ \sqrtwo i\,[\,\ulambda\,,\upsi\,]\, \ , }
in the super-multi-instanton background.
We use undertwiddling for fields in $SU(2)$ matrix notation,
${\uA} \equiv \sum_{a=1,2,3}\,A^a\tau^a/2$, where $\tau^a$ are Pauli matrices.
$A^a_{\rm inh}$ was found for the general case in Sec.~7.3 of 
Ref.~\dkmMO.

For the problem at hand we need not use the complete expression 
for  $A^a_{\rm inh}$. 
Since the action $S_{n\hbox{-}\inst}$ does not depend on the 
the supersymmetric collective coordinates for gaugino and
higgsino,  $\xi_1$ and $\xi_2$, Grassmann 
integrations over these unlifted fermionic
collective coordinates will produce a non-zero result 
only if there is an explicit dependence on $\xi_1$ and $\xi_2$ in the
integrand $A^a_{\rm inh} \ A^a_{\rm inh}$ of \Ginst.
Thus it  suffices to keep only the term in $A^a_{\rm inh}$
with explicit bilinear
dependence on $\xi_1$ and $\xi_2$, and to drop the rest. This term,
which we denote
$A^a_{\rm ss}$, is a solution of Eq.~\inhH\ with the right-hand side
made of only supersymmetric fermion zero modes for $\lambda$ and $\psi$.
The nice observation of Ref.~\FT\ is that this
bilinear piece is proportional to the ADHM field strength,\foot{This
observation
follows immediately from some of the explicit formulae in Ref.~\dkmMO; see
 Eq.~(4.3b) in the 1-instanton sector; and 
in the general $n$-instanton sector, compare the expression
for $v_{mn}$ given in Sec.~6.1, on the one hand, with the relevant
bilinear piece of the Higgs extracted from Eqs.~(7.23) and (7.5) on the
other hand.}
\eqn\Ass{ {\uA}_{\rm ss} \ = \ \sqrtwo\xi_2\sigma^{mn}\xi_1\uv^{\ADHM}_{mn}\ ,
 }
so that
\eqn\AAss{ 
 A^a_{\rm ss} \ A^a_{\rm ss} \ = \ - \xi_1^2 \xi_2^2 \ 
 v^{a \ \ADHM}_{mn} v^{a \ \ADHM}_{mn} \ . }
Substituting this into Eq.~\Ginst, performing the Grassmann integrals,
$ \int  d^2 \xi_1 \ \xi_1^2 \int d^2 \xi_2 \ \xi_2^2 \ = \ 1$,
and integrating over global translations,
\eqn\action{
\hf \, \int d^4 x_0 \   v^{a \ \ADHM}_{mn} v^{a \ \ADHM}_{mn}
   \ = \ 16\pi^2\,n\ ,}
gives the general $n$-instanton expression for  $\G(a)$ in terms of the
reduced $n$-instanton measure $d\tilde\mu_n $ of Eq.~\measuretwo :
\eqn\Ginstwo{ \G(a) \ |_{\rm n-inst} 
\ \equiv \
 {\cal G}_n\,\left({\Lambda\over a}\right)^{4n} a^2  
\ = \ 
 -16\pi^2 n \, \int\, d\tilde\mu_n \ .}
 
In what follows we derive the 
 general $n$-instanton expression for  $\F(a)$ :
\eqn\Finstwo{ \F(a) \ |_{\rm n-inst}
 \ \equiv \
 {-i \over \pi} {\cal F}_n\,\left({\Lambda\over a}\right)^{4n} a^2 
 \ = \ 
8 \pi i\ \int\, d\tilde\mu_n 
 \ . }
The advertised  all-orders relation \GnFn\
follows immediately from a comparison of \Ginstwo\ and \Finstwo.

The calculation of $\F_n$ proceeds in a few short steps:
\item{1.} Write out  $\Leff$, Eq.~\Leffdef, in components as per Ref. \dkmMO :
\eqn\Leffcomp{\eqalign{\Leff\ =\ {1\over4\pi}\,\Im\Big[
&-\F''(A)\Big(\partial_mA^\dagger\partial^mA+i\psi\delslash\psibar
+i\lambda\delslash\lambdabar+\hf(\vsd_{mn})^2\,\Big)
\cr&+{\textstyle{1\over\sqrtwo}}\F'''(A)\lambda\sigma^{mn}\psi v_{mn}+\quarter
\F''''(A)\psi^2\lambda^2\,\Big]\  ,}}
ignoring the (sub-leading) dependence on the auxiliary component
fields.
\item{2.} Concentrate on the four-fermion term in $\Leff$:
\eqn\fourf{ {-i\over32\pi}\, \F''''(a)\psi^2\lambda^2 \ , }
which gives for the 4-point Green function:
\eqn\fpoint{\eqalign{\langle\,
&\lambdabar_{\dot\alpha}(x_1)\,
\lambdabar_{\dot\beta}(x_2)\,
\psibar_{\dot\gamma}(x_3)\,
\psibar_{\dot\delta}(x_4)\, \rangle \ = \ 
\cr& {-i\over 8\pi}\, \F''''(a) \ \int d^4 y
\epsilon^{\alpha\beta}\,S_{\alpha\dalpha}(x_1-y)S_{\beta\dbeta}(x_2-y)\,
\epsilon^{\gamma\delta}\,S_{\gamma\dgamma}(x_3-y)S_{\delta\ddelta}(x_4-y)
\ ,}}
in terms of Weyl fermion propogator $S(x)_{\alpha\dalpha}$. 
\item{3.} Consider the $n$-instanton contribution to this Green function:
\eqn\Infpoint{\langle\,
\lambdabar_{\dot\alpha}(x_1)\,
\lambdabar_{\dot\beta}(x_2)\,
\psibar_{\dot\gamma}(x_3)\,
\psibar_{\dot\delta}(x_4)\, \rangle \ = \ \int\, d\mu_n \ 
\lambdabar_{\dot\alpha}(x_1)\,
\lambdabar_{\dot\beta}(x_2)\,
\psibar_{\dot\gamma}(x_3)\,
\psibar_{\dot\delta}(x_4) \ , }
where the long-distance limit of the anti-fermion components
of the multi-instanton satisfy \dkmMO :
\eqn\psibarlong{\psibar_\dalpha(x)\ =\ i\sqrtwo \ 
{\partial S_{n\hbox{-}\inst} \over \partial a}\,
\xi_1^\alpha S_{\alpha\dalpha}(x-x_0)}
and likewise
\eqn\lambdabarlong{\lambdabar_\dalpha(x)\
 =\ -i\sqrtwo  \ {\partial S_{n\hbox{-}\inst}\over \partial a}\,
 \xi_2^{\alpha} S_{\alpha\dalpha}
(x-x_0) \ .}
Here we carefully distinguish between $a$ and $\bar{a}$ and use the
fact\foot{In particular see Eq. (7.32) in the more recent,
slightly expanded version of Ref.~\dkmMO\ where $S_{n\hbox{-}\inst}$
is given explicitly in the general case of complex vev.
The corresponding expression in the old version
of ~\dkmMO\ is specific to the case of real vev
and does not make this distinction.}
that $S_{n\hbox{-}\inst}$ is at most linear in $a$ (bilinear in $a$
and $\bar a$).
\item{4.} Performing the Grassmann integrations over $ d^2 \xi_1$
and $d^2 \xi_2$ on the right hand side of \Infpoint ,
\eqn\Inptwo{
\int d^4 x_0
\epsilon^{\alpha\beta}S_{\alpha\dalpha}(x_1-x_0)S_{\beta\dbeta}(x_2-x_0)
\epsilon^{\gamma\delta}S_{\gamma\dgamma}(x_3-x_0)S_{\delta\ddelta}(x_4-x_0)
\ \int d\tilde\mu_n  \ 
\biggl( {\partial S_{n\hbox{-}\inst}\over \partial a}\biggr)^4
 }
and using the fact \dkmMO $ \ $ that the only dependence on the vev $a$ in the 
measure  
$d\tilde\mu_n $ is through $\exp ( - S_{n\hbox{-}\inst})$ and that 
$S_{n\hbox{-}\inst}$ is linear in
$a$, we finally obtain
\eqn\Inpthree{ {\partial^4 \over \partial a^4} \ \int \ d\tilde\mu_n  \ 
\int d^4 x_0
\epsilon^{\alpha\beta}\,S_{\alpha\dalpha}(x_1-x_0)S_{\beta\dbeta}(x_2-x_0)\,
\epsilon^{\gamma\delta}\,S_{
\gamma\dgamma}(x_3-x_0)S_{\delta\ddelta}(x_4-x_0) \ 
.}
\item{5.} Comparing the right hand sides of Eqs.~\fpoint\
 and \Inpthree\ we derive the desired expression \Finstwo,
and thus $\G_n=2n\F_n$.

$$\scriptstyle{************************}$$

The work of ND was supported in part by a PPARC Advanced Research
Fellowship; both ND and
 VK were supported in part by the Nuffield 
Foundation; MM was supported by the Department of Energy.

\listrefs

\bye